\documentclass[aps,pra,superscriptaddress,twocolumn,nofootinbib,floatfix,longbibliography]{revtex4-1} 

\usepackage[final]{graphicx}    
\graphicspath{{figures/}}

\usepackage{bm} 
\usepackage{amsmath}
\usepackage{amsfonts} 
\usepackage{amssymb}
\usepackage[utf8]{inputenc}
\usepackage{textcomp}
\usepackage[english]{babel}
\usepackage{hyperref}
\usepackage{soul}
\usepackage[usenames,dvipsnames]{xcolor}

\usepackage{listings}
\usepackage{color}

\definecolor{dkgreen}{rgb}{0,0.6,0}
\definecolor{gray}{rgb}{0.5,0.5,0.5}
\definecolor{mauve}{rgb}{0.58,0,0.82}

\newcommand{\ket}[1]{\left| #1 \right\rangle}
\newcommand{\bra}[1]{\left\langle #1 \right|}

\newcommand{\ketbra}[2]{| #1\rangle\!\langle #2 |}

\newcommand{\conm}[2]{\left[#1,#2\right]}
\newcommand{\aconm}[2]{\left\{#1,#2\right\}}

\begin{document}
\title{Simulation of complex dynamics of mean-field $p$-spin models using measurement-based quantum feedback control}

\author{Manuel H. Mu\~{n}oz-Arias}
\email{mhmunoz@unm.edu}
\author{Ivan H. Deutsch}
\affiliation{Center for Quantum Information and Control, CQuIC, Department of 
Physics and Astronomy, University of New Mexico, Albuquerque, New Mexico 87131, USA}
\author{Poul S. Jessen}
\affiliation{Center for Quantum Information and Control, CQuIC, College of Optical Sciences and Department of Physics, University of Arizona, Tucson, AZ 85721, USA}
\author{Pablo M. Poggi}
\affiliation{Center for Quantum Information and Control, CQuIC, Department of 
Physics and Astronomy, University of New Mexico, Albuquerque, New Mexico 87131, USA}

\begin{abstract}
We study the application of a new method for simulating nonlinear dynamics of many-body spin systems using quantum measurement and feedback [Muñoz-Arias et al., Phys. Rev. Lett. 124, 110503 (2020)] to a broad class of many-body models known as $p$-spin Hamiltonians, which describe Ising-like models on a completely connected graph with $p$-body interactions.  The method simulates the desired mean field dynamics in the thermodynamic limit by combining nonprojective measurements of a component of the collective spin with a global rotation conditioned on the measurement outcome. We apply this protocol to simulate the dynamics of the $p$-spin Hamiltonians and demonstrate how different aspects of criticality in the mean-field regime are readily accessible with our protocol.  We study applications including properties of dynamical phase transitions and the emergence of spontaneous symmetry breaking in the adiabatic dynamics of the collective spin for different values of the parameter $p$.  We also demonstrate  how this method can be employed to study the quantum-to-classical transition in the dynamics continuously as a function of system size.
\end{abstract}
\maketitle

\section{Introduction}
Using carefully manipulated quantum systems to simulate physical models of complex systems is widely seen as one of the most promising near-term applications of quantum technologies. Important advances in this direction have been demonstrated recently using trapped ions \cite{Kim2011,zhang2017_monroe,jurcevic2017,Monroe2019}, superconducting qubits \cite{xu2019,Deng2016,Salathe2015,Yanay2019}, and ultracold atoms \cite{Simon2011,Hofstetter2018,Fukuhara2013,Chen2011,Chen2010,Bloch2012}, among other platforms. For quantum simulations of many-body systems, a major goal is to be able to engineer different kinds of interactions between the constituents of the system. However, each physical platform imposes natural limitations on the type and range of such interactions, as typically seen in trapped ions with power-law decaying Ising couplings \cite{Blatt2012}, or in arrays of Rydberg atoms with the so-called kinetically constrained spin models \cite{bernien2017,ho2019}. Therefore, developing novel techniques for simulating many-body interactions is desirable and would allow quantum simulators to access a broader class of physical models.\\

In the context of quantum simulation, one tool that has been largely unexplored is measurement-based quantum feedback control (QFC) \cite{doherty2000,wiseman2009,zhang2017}, which has a long history that originated in quantum optics \cite{wiseman1994,wiseman2009}. There, one extracts information about the state of the system using (typically weak) measurements, and then uses that information to condition its future evolution. Possible applications of QFC have been studied in many different contexts over the past decades, including
deterministic generation of squeezing \cite{wiseman1994_sq,thomsen2002}, state preparation \cite{steck2004,combes2006,sayrin2011} and error suppression and correction \cite{ahn2002,vijay2012,uys2018}. The enabling power of measurements for quantum information processing has long been recognized in photonic quantum computing, where it is known that one can in principle achieve universality combining linear optics and nonGaussian measurements \cite{Kok2007}. In another application, Lloyd and Slotine \cite{Lloyd2000}, studied how QFC could be used to engineer nonlinear Schr\"odinger equations using weak measurements and feedback.

In this paper we study in detail a method proposed in \cite{munoz2019} which uses measurement-based QFC to simulate nonlinear dynamics in collective spin systems (e.g., in an ensemble of two-level atoms). In previous work we used this method to study the quantum-to-classical transition of the quantum-chaotic kicked top. Here, we study in detail the scope of this proposal and demonstrate its suitability to simulate
a broad family of spin Hamiltonians, called $p$-spin models \cite{derrida1980}. These models exhibit a broad variety of phenomena associated with nonlinear dynamics and criticality, e.g., ground state phase transitions, dynamical phase transitions, and spontaneous symmetry breaking. We show that the proposed method allows us to probe these features close to the thermodynamic limit in the mean-field regime. 

The remainder of this paper is organized as follows. In Sec. \ref{sec:simul_method} we present an overview of the method originally described in \cite{munoz2019}, and discuss the class of conditioned unitary operations which are most suitable to simulate Hamiltonian dynamics. We also illustrate the existence of measurement conditions which are optimal to achieve such simulation, and compare our formalism with the theory of continuous measurements and Markovian feedback. Then, in Sec. \ref{subsec:p_spin_models}, we 
introduce a summary of the most important features of $p$-spin models, focusing on how phase transitions of different orders are obtained when the interaction degree $p$ is changed. In Sec. \ref{subsec:p_spin_simu} and \ref{subsec:optimal_msmt} we show how to apply this method to simulate the dynamics of these models in the mean-field regime, and derive analytically the  measurement strength regime which optimizes the success of the method. We then present a series of applications of our feedback simulation of the $p$-spin Hamiltonians. In Sec. \ref{subsec:phase_spaces} we study the corresponding classical phase space structures and discuss how well they can be resolved. In Sec. \ref{subsec:dpt} we demonstrate how different signatures of dynamical phase transitions are readily accessed with this protocol. In Sec. \ref{subsec:symmetry_breaking} we illustrate the emergence of spontaneous symmetry breaking in the dynamics, induced by the measurements performed on the system. Then, in Sec. \ref{subsec:quantum_classical} we study in detail how well the simulation is achieved in the thermodynamic limit as the number of particles is increased. Finally, in Sec. \ref{sec:future} we summarize and discuss other potential applications of the proposed method.

\section{Simulation via quantum measurement and feedback}
\label{sec:simul_method}

\subsection{Overview of the method}
\label{sec:overview}
\begin{figure}[!t]
 \centering{\includegraphics[width=0.48\textwidth]{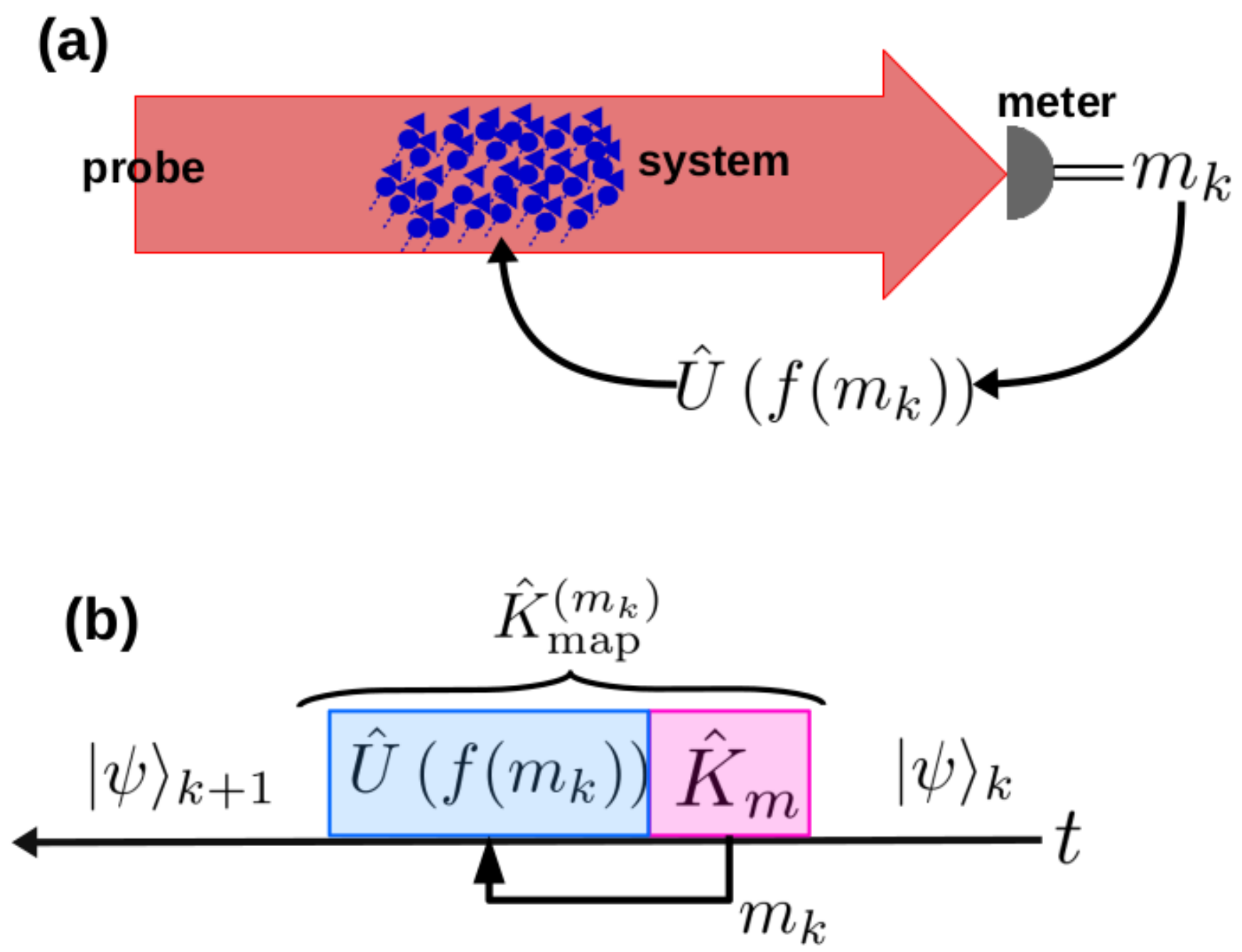}}
\caption{Schematic representation of the quantum feedback control (QFC) simulation scheme \textbf{(a)} At each simulation step, the system interacts with a probe which is then measured projectively. The measurement outcome $m_k$ is used to condition a unitary map $\hat{U}$ which then acts on the original system.  \textbf{(b)} Stroboscopic time evolution of the quantum state for the $k$-th protocol step. The state is updated via quantum Bayes' rule as in Eq. (\ref{eqn:map_quantum_bayes}), giving a map composed of a nonprojective measurement of the type described in Eq. (\ref{eqn:kraus_ope}) followed by a unitary map conditioned on the measurement outcome.}\label{fig:protocol_sketch}
\end{figure}
We summarize here the simulation protocol originally proposed in \cite{munoz2019}. We consider an ensemble of $N$ spin-$\frac{1}{2}$ particles described by collective spin operators $(\hat{J}_x,\hat{J}_y,\hat{J}_z)$, where
\begin{equation}
\label{eqn:collective_operators}
    \hat{J_\alpha}=\frac{1}{2}\sum\limits_{i=1}^N \hat{\sigma}^{(\alpha)}_i
\end{equation}
\noindent and $\hat{\sigma}^{(\alpha)}_i$ with $\alpha=x,y,z$ is a  Pauli 
operator corresponding to the $i$-th particle. At the initial time $t_0$, we 
assume the state of the system to be given by a spin coherent state (SCS)   
$\ket{\psi_0}=\ket{\uparrow_{\vec{n}}}^{\otimes N}=\ket{\theta,\phi}$, where all 
particles are polarized along a particular direction on the unit sphere 
$\vec{n}$, specified by angles $(\theta,\phi)$ on the sphere. The system then 
evolves in discrete time steps $\{t_k\}$ where the state is described by $\{ 
\ket{\psi_k} \}$, $k=0,1,\ldots,N_{steps}$. As depicted in Fig. 
\ref{fig:protocol_sketch}, at each time $t=t_k$, a nonprojective measurement of 
a component of the collective spin (say $\hat{J}_z$) is performed, yielding a 
measurement outcome $m_k$. The probability of seeing a particular measurement 
outcome $m_k$ is given by the Born rule
\begin{equation}
    P(m_k)=\bra{\psi_k}\hat{K}^\dagger_{m_k}\hat{K}_{m_k}\ket{\psi_k}
\end{equation}
\noindent where $\hat{K}_{m_k}$ is the Kraus operator describing the nonprojective measurement which has the form \cite{Jacobs2014}
\begin{equation}
\label{eqn:kraus_ope}
\hat{K}_{m_k}=\frac{1}{(2\pi\sigma^2)^{1/4}}e^{-\frac{1}{4\sigma^2}\left(\hat{J}_z - m_k\right)^2}
\end{equation}

\noindent where $\sigma$ is the measurement resolution. After the measurement, a unitary operation $\hat{U}\left(f(m_k)\right)$ is applied to the system. This operation is conditioned on the measurement outcome via a \textit{feedback policy} $f(m)$, which can be an arbitrary function of $m$. The unitary map $\hat{U}$ will typically be constrained to a restricted set of operations which can be easily implemented. An example would be global SU(2) rotations for collective spin systems \cite{montano2018}. At the end of the $k$-th time step the state of the system is updated following quantum Bayes rule \cite{schack2001}, thus leading to the following map,
\begin{equation}
\label{eqn:map_quantum_bayes}
\ket{\psi_{k+1}}=\frac{1}{\sqrt{P(m_k)}}\hat{U}(f(m_k))\hat{K}_{m_k}\ket{\psi_k}.
\end{equation}

In the following we will be interested in the discrete quantum trajectory $\{\ket{\psi_k}\}$ generated by this protocol, which is conditioned on a set of measurement outcomes $\{m_k\}$.

\subsection{Choice of conditioned unitary and optimal measurement strength}

\label{subsec:control_hamil}
The protocol described in the previous subsection leads to a broad class of dynamics, which in particular encompasses known applications of quantum feedback control \cite{zhang2017}. In the following we show how a particular choice of the unitary $\hat{U}(f(m))$ leads naturally to an effective dynamics which \textit{simulates} a desired Hamiltonian. In order to motivate this argument, consider the limit of infinitesimally short time steps, with length $t_{k+1} - t_k\rightarrow dt$. In this case the dynamics is described by the theory of continuous weak measurements \cite{wiseman2009,Jacobs2014}. The measurement record is now a continuous function of time $\{m_k\}\rightarrow \mathcal{M}(t)$ whose evolution is described by the equation
\begin{equation}
    \mathcal{M}(t)dt=\langle \hat{A}\rangle dt + \frac{1}{\sqrt{\kappa}}dW
    \label{ec:mtmt_record}
\end{equation}
\noindent where $\langle \hat{A}\rangle$ is computed over the state $\hat{\rho}(t)=\ketbra{\psi(t)}{\psi(t)}$, $\kappa\:dt=\sigma^{-2}$ is the inverse of the squared measurement strength and $dW$ is a Wiener increment \cite{Jacobs2010,Jacobs2014}. While in our case $\hat{A}=\hat{J}_z$, the argument presented here is general. 

Wiseman and Milburn showed \cite{wiseman1994,wiseman2009} that a system undergoing continuous weak measurements of the observable $\hat{A}$ and driven by a control Hamiltonian of the form
\begin{equation}
    \hat{H}(t) = \lambda \mathcal{M}(t) \hat{F},
    \label{ec:hami_fb}
\end{equation}

\noindent with $\lambda$ the feedback strength and $\hat{F}$ an hermitian operator, obeys the following QFC master equation
\begin{widetext}
\begin{equation}
d\hat{\rho} = -i\conm{ \frac{\lambda}{4}\aconm{\hat{A}}{\hat{F}}}{\hat{\rho}}dt + \frac{\kappa}{4} \mathcal{D}\left[\hat{A}-i\frac{2\lambda}{\kappa}\hat{F}\right]\hat{\rho}\:dt + \sqrt{\frac{\kappa}{4}}\mathcal{H}\left[\hat{A}-i\frac{2\lambda}{\kappa}\hat{F}\right]\hat{\rho}\: dW,
\label{ec:me_feedback}
\end{equation}
\end{widetext}
where we have assumed perfect measurement efficiency and defined the superoperators 
\begin{eqnarray}
\mathcal{D}[\hat{c}]\hat{\rho} &=& -\frac{1}{2}\left(\hat{c}^\dagger \hat{c} \hat{\rho} + \hat{\rho} \hat{c}^\dagger \hat{c} - 2\hat{c} \hat{\rho} \hat{c}^\dagger\right)\\ 
\mathcal{H}[\hat{c}]\rho &=& \hat{c}\hat{\rho} + \hat{\rho} \hat{c}^\dagger - \mathrm{Tr}\left[(\hat{c}+\hat{c}^\dagger)\hat{\rho}\right].
\end{eqnarray}
Eq. (\ref{ec:me_feedback}) describes the general dynamics of the system for an arbitrary choice of the feedback operator $\hat{F}$. Note that
the overall effect of the feedback in Eq. (\ref{ec:me_feedback}) has both unitary and nonunitary contributions, the latter being particularly important in some applications of QFC, such as state stabilization. However, if we choose the feedback Hamiltonian to coincide (up to a multiplicative constant) with the operator being monitored, i.e.  $\hat{F}=\hat{A}$, Eq. (\ref{ec:me_feedback}) reduces to 

\begin{widetext}

\begin{equation}
d\hat{\rho} = -i\conm{ \frac{\lambda}{2}\hat{A}^2}{\hat{\rho}}dt - \frac{\gamma}{2} [\hat{A},[\hat{A},\hat{\rho}]]dt
+ \sqrt{\frac{\kappa}{4}} \left( \aconm{\hat{A}}{\hat{\rho}}-2\langle  \hat{A}\rangle -i\frac{2\lambda}{\kappa} \conm{\hat{A}}{\hat{\rho}} \right) dW,
\label{ec:me_feedback_sim}
\end{equation}
\end{widetext}

Eq. (\ref{ec:me_feedback_sim}) describes the evolution of a quantum system which is: i) driven by a Hamiltonian $\hat{H}_{FB}=\frac{\lambda}{2}\hat{A}^2$, ii) subjected to dephasing in the basis of $\hat{A}$ with a rate

\begin{equation}
    \gamma = \frac{\kappa}{4} + \frac{\lambda^2}{\kappa},
    \label{ec:dephasing_rate}
\end{equation}

\noindent and iii) driven by a stochastic term which appears as proportional to $dW$ in the equation and is zero if we average over measurement outcomes \cite{Jacobs2014}. Focusing on (i) and (ii), we readily see that choosing $\hat{F}=\hat{A}$ decouples the deterministic effect of the feedback into a completely unitary part (i) and a nonunitary contribution (ii) which adds to the unavoidable dephasing induced by the measurement. This general argument reveals the existence of a regime under which measurement-based feedback control \textit{simulates} unitary evolution generated by a Hamiltonian $\hat{H}_{FB}$ at the expense of additional dephasing. Note also that the total dephasing rate, Eq.  (\ref{ec:dephasing_rate}), is not a monotonic function of the measurement rate $\kappa$. This can be understood from the fact that  $\kappa\rightarrow \infty$ (and conversely $\sigma \rightarrow 0$) is the limit of \textit{projective} measurements, in which a very accurate estimate of $\hat{A}$ is obtained at the expense of a large measurement backaction on the system, which dominates over the unitary evolution leading to $\gamma\rightarrow\infty$. The opposite limit $\kappa\rightarrow 0$ (and conversely $\sigma \rightarrow \infty$) is that of \textit{weak} measurements. There, the measurement disturbs the state of the system only slightly, but obtains a very inaccurate estimation of $\hat{A}$. As a consequence, the control Hamiltonian in Eq. (\ref{ec:hami_fb}) feeds back mostly noise into the system, thus leading again to $\gamma\rightarrow\infty$. It is then clear that a minimum value of $\gamma$ is achieved by an intermediate measurement rate $\kappa_{opt}$, which demonstrates the existence of an information gain - disturbance tradeoff in the dynamics simulated by the feedback procedure \cite{munoz2019}. Exploiting this fact is an integral part of our proposal, and so we will be interested in working at the point of optimal measurement in all cases.

\subsection{Relation to continuous measurement and Markovian feedback}

Although in the previous section we have motivated our choice of control Hamiltonian using the well-established theory of continuous weak measurements, it is important to point out that the general (discrete time) simulation protocol layed out in Sec. \ref{sec:overview} actually allows for a more general class of dynamics which are not described by the stochastic master equation in Eq. (\ref{ec:me_feedback}). This is because the control Hamiltonian in Eq. (\ref{ec:hami_fb}) is restricted to be a linear function of the measurement record $\mathcal{M}(t)$ in the continuous case, as higher-order powers are ill-defined \cite{wiseman1994_sq}. To see why this is the case, consider a modified control Hamiltonian $\hat{H}_n(t)=\lambda \mathcal{M}(t)^n\hat{F}$ with $n$ a positive integer and recall from Eq. (\ref{ec:mtmt_record}) that $\mathcal{M}(t) = \langle \hat{A}\rangle + \frac{1}{\sqrt{\kappa}}\frac{dW}{dt}$. Then, the change induced in the state by this Hamiltonian is
\begin{equation}
    d\ket{\psi}=-i\lambda \mathcal{M}(t)^n\hat{F}\ket{\psi}dt \sim -i\lambda \left(\frac{dW}{dt}\right)^n\hat{F}\ket{\psi}dt,
\end{equation}
\noindent the last term being the leading order in $dt$. Recalling that $dW^2=dt$, it is then clear that in order to keep $d\ket{\psi}\rightarrow 0$ as $dt\rightarrow 0$, the exponent $n$ cannot be higher than 1. For $n=2$, actually, this term remains $\mathcal{O}(1)$, independent of $dt$. \\

In contrast, in the discrete time case the feedback policy can be any (nonlinear) function of the measurement outcome $m$, which gives us access to different classes of dynamics, as we will illustrate in the next section. Note also that, as shown in \cite{munoz2019}, the measurement outcome $m$ can be taken as the time-averaged measurement record $\mathcal{M}(t)$, which makes the discrete time evolution essentially non-Markovian. Of course, the downside of the discrete time formalism is that we cannot use the powerful tools of stochastic calculus. However, as we will show in the next section, our approach allows for a useful analytical treatment when working in the regime of large system size $N=2J \gg 1$, by employing the Holstein-Primakoff approximation~\cite{Holstein1940}. This is the regime of greatest interest for simulation of mean-field dynamics, as we will see below.

\section{$p$-spin systems}
The $p$-spin models are a family of simple models describing the competition between two magnetically distinct orderings in an ensemble of $N$ spin-$1/2$ particles. The ensemble is subjected to an external uniform magnetic field inducing paramagnetic order, and an infinite range $p$-body Ising-like interaction inducing ferromagnetic order. In this family of models one recovers for $p=2$ the well known Curie-Weiss ferromagnet~\cite{Chayes2008,Krzakala2008,Kochmanski2013}, which is the Lipkin-Meshkov-Glick (LMG) model when interactions are infinite range ~\cite{Lipkin1965}. For general $p$, the interplay of these two incompatible orderings gives rise to rich critical phenomena. We parametrize the $p$-spin Hamiltonian as a dimensionless universal Hamiltonian
\begin{equation}
 \label{eqn:p_spin_hamil}
 \hat{H}_p = -(1-s)\hat{J}_y - \frac{s}{p J^{p-1}}\hat{J}_z^p,
\end{equation}
where we have chosen the external field to be along the $y$-axis and the interactions to be of $z$-type, $s\in[0,1]$ is an interpolation parameter controlling the degree of mixture between the two distinct orderings, and $\hat{J}_{\alpha}$ are collective spin operators as in Eq. (\ref{eqn:collective_operators}). Our choice of parametrization is in natural units, and naturally generalizes the mean-field dynamics of the LMG model to models with higher $p$. 

\subsection{Summary of properties of $p$-spin models}
\label{subsec:p_spin_models}
Since the Hamiltonian preserves the total spin, that is, $[\hat{H}_p, \hat{\vec{J}}^2]=0$, the dynamics of the system is constrained to the symmetric subspace of the ensemble of spin-$1/2$ particles, spanned by $N+1$ Dicke states $\{|J,J\rangle, |J,J-1\rangle,...,|J,-J\rangle\}$, which correspond to permutation symmetric states. A salient feature of these models is the existence of phase transitions of varying order depending on $p$. For $p=2$, the transition is second order (continuous), while for $p>2$, it is first order (discontinuous). We refer to~\cite{Filippone2011} for a thorough study of quantum phase transitions for a general model of collective interacting spins. In the quantum regime, the study of the behavior of the spectral gap in these systems has received special attention in recent years, and it was shown that for $p=2$ the gap closes polynomialy with $N$ and for $p>2$ it closes exponentially~\cite{Jorg2010}. The latter has strong consequences for dynamical processes such as adiabatic quantum computation~\cite{Kong2017} and quantum annealing~\cite{Matsuura2017} where $p$-spin models have been studied as systems constituting hard problems for annealers to solve~\cite{Jorg2010}.

\begin{figure}[!t]
 \centering{\includegraphics[width=0.48\textwidth]{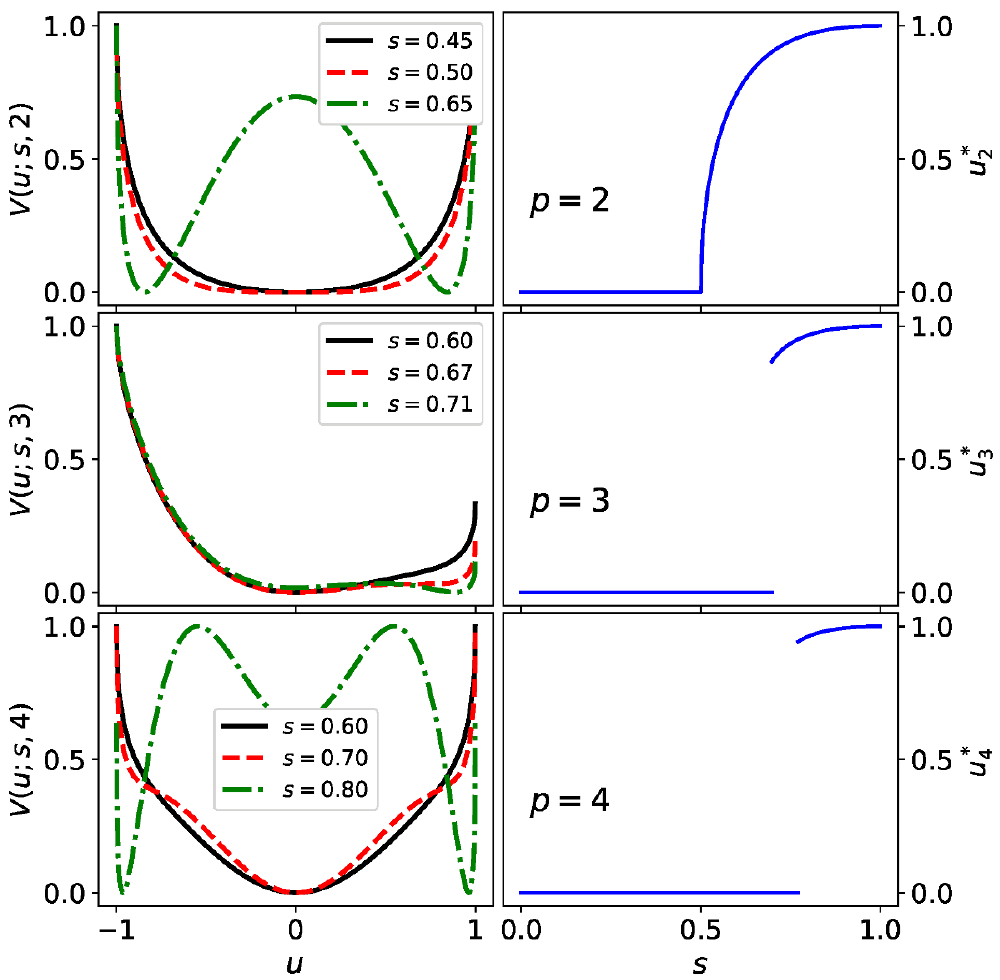}}
\caption{Equilibrium phase transitions in the $p$-spin model. (\textbf{left}) Semiclassical energy function of Eq. (\ref{eqn:pseudo_potential}) as a 
function of the direction of the collective spin vector, parametrized by $u=\cos(\theta)$ with $\phi=\pi/2$, for three different values of $s$. 
From top to bottom we show the cases of $p=2,3,4$, 
respectively. We normalized each curve by the difference between its maximum and minimum values, so that they lie within the interval $(0,1)$. 
(\textbf{right}) Global minimum of the 
pseudo potential as a function of $s$. Notice the 
difference in the continuity of the curves between 
the cases $p=2$ (second order phase transition) and $p\ge3$ (first order phase transition).}\label{fig:pseudo_potential}
\end{figure}

Properties of the Hamiltonian in Eq. (\ref{eqn:p_spin_hamil}) in the mean-field case can be obtained by studying the semiclassical energy function in the thermodynamic limit. We construct this energy function by taking the expected value of the Hamiltonian in Eq.  (\ref{eqn:p_spin_hamil}) in a spin coherent state, neglecting correlations and, in the limit $J \to \infty$,  defining the classical variables $\vec{X} = \langle \hat{\vec{J}} \rangle/{J}$. After this procedure one finds
\begin{equation}
 \label{eqn:pseudo_potential}
 V(u,\phi;s,p) = -(1-s)\sqrt{1-u^2}\sin(\phi) - \frac{s}{p}u^p, 
\end{equation}
where we have written the classical variables $\vec{X}$ in spherical coordinates $(X,Y,Z)=(\sin(\theta)\cos(\phi), \sin(\theta)\sin(\phi), \cos(\theta))$, and defined $u = \cos(\theta)$. In these models, phase transitions can be studied by analyzing an order parameter (OP) in the ground state. Here we take the OP to be the magnetization along the $z$-direction. In the thermodynamic limit, we can obtain the value of the ground state OP for a given value of $s$ as the value of $u$ at which Eq. 
(\ref{eqn:pseudo_potential}) has its global minimum. It is straightforward to see that the minimum condition implies $\phi=\pi/2$. As a consequence, the identification of the extreme values of $V(u,\phi=\pi/2)\equiv V(u)$ is the main tool to study phase transitions in our model. It is easy to check that $u=0$ is always an extreme value of $V$ and new extrema appear when 
\begin{equation}
 \label{eqn:new_minima_equation}
 u^{2(p-1)} - u^{2(p-2)} + \left( \frac{1-s}{s} \right)^2 = 0,
\end{equation}
has real-valued solutions. For $s=0$ the global minimum is at $u=0$. As $s$ is increased other extreme values appear, and eventually one of them becomes the new global minimum. When we observe either a nonanaliticity or a discontinuous jump in the OP, this indicates that the system has moved to a new and more stable energy configuration and thus a phase transition has taken place. In the left column of Fig. \ref{fig:pseudo_potential} we present examples of semiclassical energy curves for systems with $p=2,3,4$ (top to bottom, respectively). A marked difference in the position of the global minimum can be seen between the two curves in red and black and the green one. For the latter, the global minimum is located at a value of $u\ne0$.

Consider now a real-valued solution of Eq. (\ref{eqn:new_minima_equation}), say 
$\tilde{u}_p$. If there is a value of $s$ for which $\tilde{u}_p$ becomes the position of the global minimum, then the algebraic inequality 
\begin{equation}
\label{eqn:global_minima_eq}
 V(\tilde{u}_p; s,p) \le -(1-s),
\end{equation}
saturates at that value of $s$. Such value of $s$ marks the boundary between two different phases and we will referred to it as the equilibrium critical point,
$s=s_c^{\rm eq}$.
Let us now study Eq. (\ref{eqn:new_minima_equation}) and Eq. 
(\ref{eqn:global_minima_eq}) for the three systems with $p=2,3,4$. When $p=2$ one can show  that new extreme values of $V(u;s,p)$ exist when $s\ge0.5$ and they have the form 
\begin{equation}
 u(s) = \sqrt{1-\left(\frac{1-s}{s}\right)^2},
\end{equation}
where we have have only considered the positive 
branch. This new extreme value becomes the global minimum for values of $s$ 
such that $(s-1/2)^2 \ge 0$, which is saturated when $s=0.5$, hence $s_c^{\rm (eq)} = 0.5$. Therefore, 
for $p=2$ the value of $s$ at which new extreme values appear and at which 
they become the global minimum coincide, the phase transition is a continuous, second order one. This fact is seen by looking at the OP curve as a function of the control parameter $s$, shown in top-right panel of Fig. \ref{fig:pseudo_potential}.

In the case of $p=3$ new extreme values appear when $s\ge2/3$ and they have 
the form
\begin{equation}
 u(s) = \sqrt{\frac{1}{2} + \frac{1}{2}\sqrt{1-4\left(\frac{1-s}{s}\right)^2}}.
\end{equation}
This new extreme value becomes the global minimum at $s_c^{\rm (eq)}=0.697831$. For $p=4$ a 
similar situation happens, new extreme values exist when  $s\ge\frac{3}{23}(9-2\sqrt{3})\approx0.722073$, and they become the global minimum at $s_c^{\rm (eq)} = 0.771429$. Notice how for both $p=3$ and $p=4$, the emergence  of new extreme values and their transition to be the global minimum do not occur at the same value of $s$. Thus, we observe a discontinuous magnetization (featuring an abrupt jump) as a function of $s$, as can be seen in the middle and bottom panels in Fig. \ref{fig:pseudo_potential}. This behavior is indicative of a first order phase transition.

\subsection{Feedback protocol for simulating $p$-spin dynamics}
\label{subsec:p_spin_simu}
For the case $p=2$, the mean-field dynamics of a Hamiltonian proportional to $\hat{J}_z^2$ is obtained by linearizing the quantum fluctuations around the mean, in which case $\hat{J}_z^2\to 2\langle\hat{J}_z\rangle\hat{J}_z$. In the QFC protocol in~\cite{munoz2019} the measurement outcome $m$ provides an estimate of $\langle\hat{J}_z\rangle$, in the limit in which the signal dominates over the noise. We thus simulate the mean-field dynamics of the quadratic ``twisting Hamiltonian'' through a feedback policy that induces a collective rotation generated by the term $m\hat{J_z}$.
In order to simulate the dynamics of the $p$-spin model we follow a similar argument. In the mean-field limit we can write the interaction term in Eq. (\ref{eqn:p_spin_hamil}) as $\hat{J}_z^p\to p\langle\hat{J}_z\rangle^{p-1}\hat{J}_z$, and thus we obtain the mean-field Hamiltonian
\begin{equation}
\label{eqn:p_spin_hamil_mf}
\hat{H}_{p}^{\rm(mf)} = -(1-s)\hat{J}_y -\frac{s}{J^{p-1}}\langle\hat{J}_z\rangle^{p-1}\hat{J}_z.
\end{equation}
Note that in the mean-field limit, the interaction term is seen as a rotation around the $z$-axis by an angle proportional to the $(p-1)$-power of the expected value of $\hat{J}_z$. Since the measurement outcome $m$ provides information about the desired expected value, we choose the feedback unitary map 
\begin{equation}
\label{eqn:p_spin_feed_policy}
\hat{U}\left(f(m)\right) = {\rm exp}\left(i\Delta t(1-s)\hat{J}_y + i\Delta t\frac{s m^{p-1}}{J^{p-1}}\hat{J_z}\right),
\end{equation}
to simulate the desired dynamics over a time step $\Delta t$.

\subsection{Derivation of the optimal measurement regime}
\label{subsec:optimal_msmt}
As discussed in Section \ref{subsec:control_hamil}, we wish to operate this simulation scheme at the optimal value of the measurement resolution $\sigma$. In order to explicitly find such optimal $\sigma$, we write the map evolving the normalized vector of expectation values,  $\vec{X} = \langle \hat{\vec{J}} \rangle/J$, after the action of a single protocol step, and determine the measurement strength that best approximates the mean field dynamics. In the limit of a large spin ensemble, $N\gg1$, we can use that map to write down an analytic expression for the optimal value of $\sigma$. To work in this limit it is convenient  use the co-moving Holstein-Primakoff (H-P) approximation~\cite{Holstein1940}, the details of which were laid out in~\cite{munoz2019}, In the limit $N\gg1$, to a good approximation, we can write the state of the system as a Gaussian state. Thus, the vector of expectation values $\langle\hat{\vec{J}}\rangle$ and the symmetric covariance matrix $\mathbb{V}_{\gamma\nu} = \frac{1}{2}\left(\langle\{\hat{J}_\gamma , \hat{J}_\nu\}\rangle - 2\langle\hat{J}_\gamma\rangle\langle\hat{J}_\nu\rangle \right)$ completely determine the state of the system. In this limit the H-P approximation consists of mapping an initial spin coherent state to the vacuum of a bosonic mode on the tangent plane to the sphere at the position of the mean spin vector. Then, we construct quadrature operators on the local basis of the plane out of the collective spin operators. In the H-P plane, one can easily compute the action of $\hat{K}_m$ on the state (explicit expressions are given in~\cite{munoz2019}). Using the H-P approximation we can recast a single step of the protocol as consisting of the following parts. First, we change the basis from space-fixed Cartesian coordinates to the local basis on the plane, which is achieved by a rotation matrix $\mathbb{A}$\footnote{This is the same rotation matrix connecting Cartesian coordinates with spherical coordinates, except that the spherical basis is ordered as $(\vec{e}_{\phi},-\vec{e}_{\theta},\vec{e}_r)$}. Next we update the entries of $\vec{X}$ and $\mathbb{V}$ under the action of the measurement. Then, after rotating back to the original coordinates, we apply the rotations specified by $\hat{U}\left(f(m)\right)$. 

This calculation simplifies considerably if we write the feedback unitary map in Eq. (\ref{eqn:p_spin_feed_policy}) as a series of rotations around the axes $x$ and $z$. This operator then takes the form
\begin{equation}
\label{eqn:p_spin_feed_policy_canonical}
\hat{U}\left(f(m)\right) = e^{i\left(\alpha\hat{J}_y + \beta\hat{J}_z\right)}=e^{i\varphi\hat{J}_x}e^{-\gamma\hat{J}_z}e^{-i\varphi\hat{J}_x},
\end{equation}
where $\alpha = \Delta t(1-s)$, $\beta=\Delta t s \frac{m^{p-1}}{J^{p-1}}$ and $\gamma=\sqrt{\alpha^2 + \beta^2}$, $\varphi = \sin^{-1}(\frac{\alpha}{\gamma})$. 

Using the H-P approximation and this last expression we computed the explicit form of the map evolving $\vec{X}$ after one protocol step, yielding
\begin{widetext}
\begin{eqnarray}
\label{eqn:explicit_one_step_evo}
X_{i+1} &=& \left(  1-\eta_1\mathbb{V}_{22}Z_i \right) \left[ \cos(\gamma)X_i - \cos(\varphi)Y_i \right] - \mathbb{V}_{12}\eta_1\left[ \cos(\varphi)X_i +\cos{\gamma}Y_i \right] + \left( Z_i + \eta_1\mathbb{V}_{22} \right)\sin(\gamma)\cos(\varphi), \nonumber \\
Y_{i+1} &=& \left(  1-\eta_1\mathbb{V}_{22}Z_i \right) \left[ \cos(\varphi)\sin(\gamma)X_i + (\cos(\gamma)\cos^2(\varphi)+\sin^2(\varphi))Y_i \right] \nonumber \\ 
&+& \mathbb{V}_{12}\eta_1\left[ (\cos(\gamma)\cos^2(\varphi) + \sin^2(\varphi))X_i - \cos(\varphi)\sin(\gamma)Y_i \right]  + \left( Z_i + \mathbb{V}_{22}\eta_1(1-Z_i^2)\cos(\varphi)\sin(\varphi)(1-\cos(\gamma)) \right), \\
Z_{i+1} &=& \left(  1-\eta_1\mathbb{V}_{22}Z_i \right)  \left[ -\sin(\gamma)\sin(\varphi)X_i + \cos(\varphi)\sin(\varphi)(1-\cos(\gamma))Y_i \right] \nonumber \\ 
&+& \mathbb{V}_{12}\eta_1\left[ \cos(\varphi)\sin(\varphi)(1-\cos(\gamma))X_i + \sin(\gamma)\sin(\varphi)Y_i \right] + \left( Z_i + \mathbb{V}_{22}\eta_1(1-Z_i^2)\right)\left( \cos^2(\varphi) + \cos(\gamma)\sin^2(\varphi) \right), \nonumber
\end{eqnarray}
\end{widetext}
where $\eta_1$ and $\eta_2$ are two normally distributed random variables given by 
\begin{eqnarray}
\label{eqn:random_vars}
\eta_1 &=& \frac{m_\theta}{\sigma^2} \equiv \mathcal{N}(0,\sigma_1^2),\enspace\text{with}\enspace \sigma_1^2 = \frac{\sigma^2 + (\Delta\hat{J}_z^2)_k}{\sigma^4}, \\
\eta_2 &=& \frac{m_{\theta}}{J/W} \equiv \mathcal{N}(0, \sigma_2^2),\enspace\text{with}\enspace \sigma_2^2 = \frac{W^2}{J^2}\left(\sigma^2 + (\Delta\hat{J}_z^2)_k \right), \nonumber
\end{eqnarray}
representing the randomness coming from the noisy measurement and an imperfect feedback operation, respectively. We will refer to the measurement noise as ``shot noise'' as in a physical implementation with a laser probe. Here we have defined $m_\theta = m - \langle\hat{J}_z\rangle$ and $W=(\Delta t s)^{\frac{1}{p-1}}$. Also, $(\Delta \hat{J}_z^2)_k$ is the spin uncertainty which we refer to as ``projection noise'' of the state at the $k$-th evolution step. Note that we also obtain an explicit map for the evolution of the covariance matrix $\mathbb{V}$, which is not shown here. We point out that $\eta_2$ does not appear explicitly in Eq. (\ref{eqn:explicit_one_step_evo}), however it is present in the argument of the trigonometric functions via the parameter $\beta$, since $\beta = (WZ + \eta_2)^{p-1}$.\\

A Taylor expansion of the trigonometric functions and square roots in Eq. (\ref{eqn:explicit_one_step_evo}) shows that the first nonvanishing terms are linear in both $\eta_1$ and $\eta_2$. Thus, finding the optimal value of the measurement resolution requires the minimization of a convex combination of the two noise variances. Both random variables are centered at zero, and thus we minimize
\begin{equation}
\label{eqn:optimal_func}
f(\sigma_1^2, \sigma_2^2) = \sigma_1^2 + \sigma_2^2.
\end{equation}
If the time-evolved state remains, at all times, close to a spin coherent state, then we can consider $(\Delta\hat{J}_z^2)_k \sim J/2$, and we parametrize the measurement resolution as proportional to the projection noise, $\sigma = \mu\sqrt{J}$. With these two definitions one can easily find the value of $\mu$ minimizing $f(\sigma_1^2, \sigma_2^2)$, yielding
\begin{equation}
\label{eqn:optimal_mu}
\mu_{\rm opt} = \frac{1}{2(\Delta t s)^{\frac{1}{p-1}}}\sqrt{1 + \sqrt{1 + (\Delta t s)^{\frac{2}{p-1}}}}.
\end{equation}

\begin{figure}[t!]
 \centering{\includegraphics[width=0.31\textwidth]{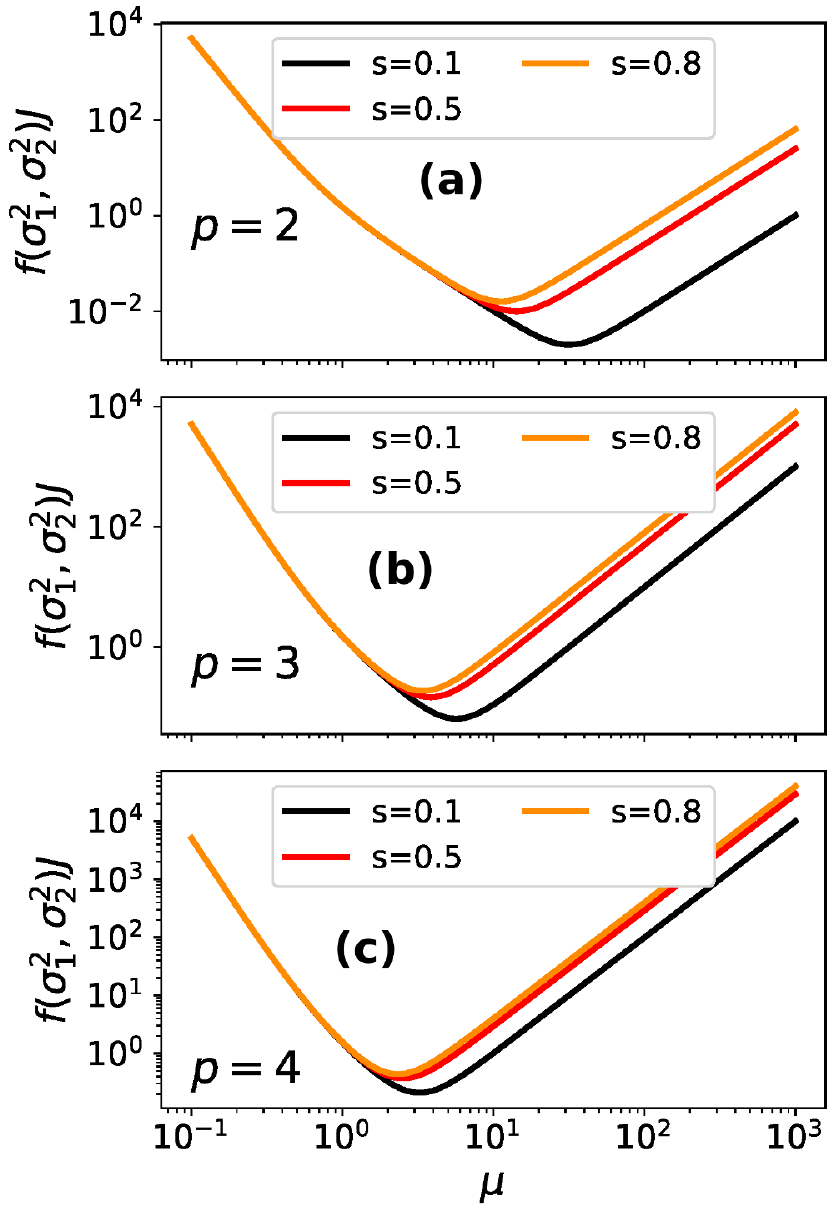}}
\caption{Analysis of the optimal measurement strength for the simulation scheme. In each panel the normalized convex combination of the variances of the two random variables $\eta_1$ and $\eta_2$ appearing in Eq. (\ref{eqn:explicit_one_step_evo}) are plotted as a function of the measurement strength parameter $\mu=\sigma/\sqrt{J}$. The case $p=2$ is shown in  \textbf{(a)}, $p=3$ in \textbf{(b)}, and $p=4$ in \textbf{(c)}. In all cases we observe the existence of an optimal (minimum) value for the measurement resolution. As $p$ increases, curves for different values of $s$ converge.}\label{fig:optimal}
\end{figure}

We study the behavior of the function $f(\sigma_1^2, \sigma_2^2)$ in a wide range of values of $\mu$ in Fig. \ref{fig:optimal}, from which several features manifest. First, the existence of a minimum is evident in all curves, at a value of $\mu_{\rm opt}\sim 1$. This is expected, since a strong measurement would lead to excessive measurement backaction, and a weak measurement would not extract sufficient information for useful feedback. The optimum $\mu$ gives the best balance to this tradeoff. We notice the optimal value moves towards smaller values of $\mu$ as $s$ and/or $p$ increases. However, for larger $p$ the position of the minimum becomes increasingly insensitive to the value of $s$; see Fig. \ref{fig:optimal}a and Fig. \ref{fig:optimal}c. Finally, for larger $p$, we observe narrower curves, indicating that the simulated dynamics is less robust to deviations in $\mu$ than for $p=2$. This last fact already hints to a relation between the value of $p$ and the ease of simulating the respective mean-field $p$-spin dynamics. We will explore this relation from different points of view with the applications presented in the next section.
\section{Applications}

\subsection{Constructing phase-space portraits}
\label{subsec:phase_spaces}
\begin{figure*}[!ht]
 \centering{\includegraphics[width=0.99\textwidth]{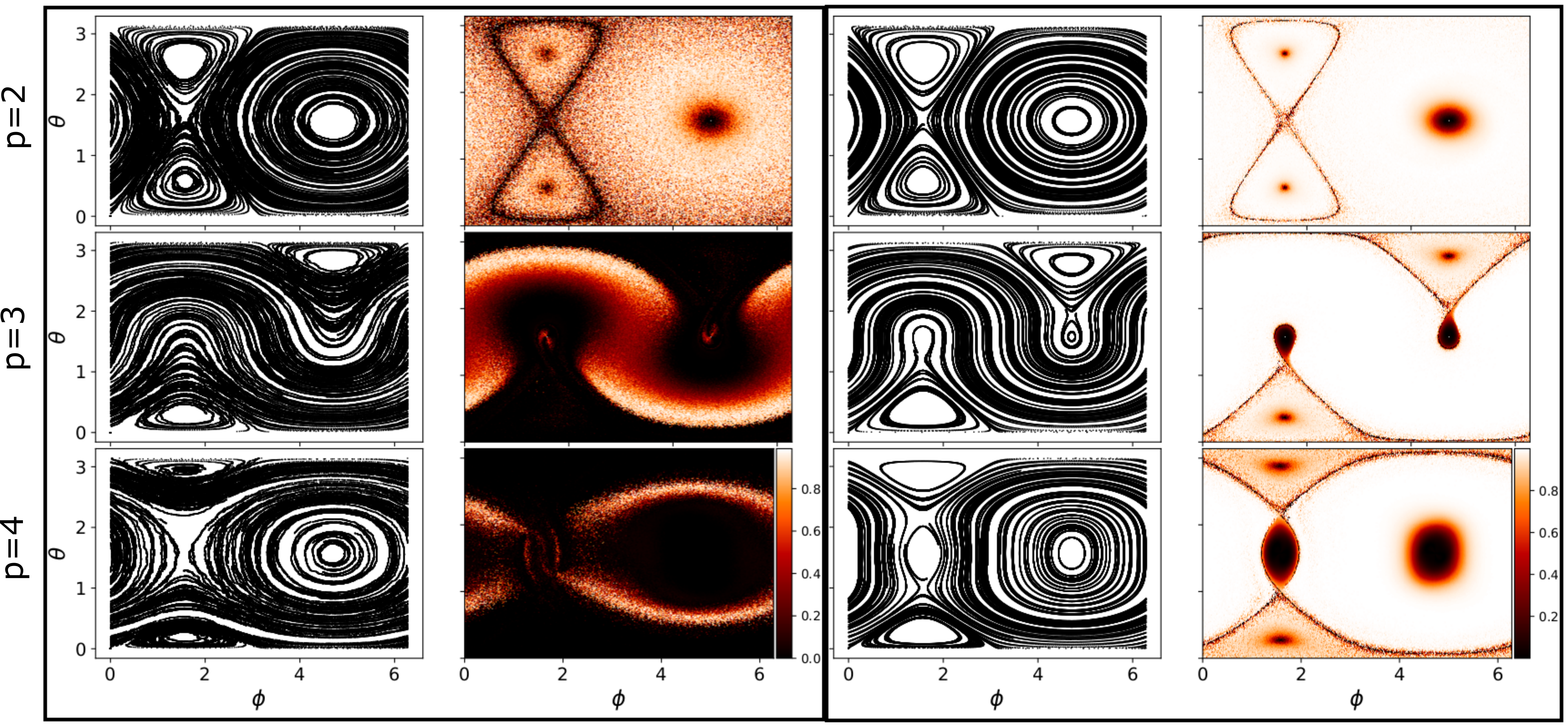}}
\caption{Phase spaces and point to point similarities for the simulation of the mean-field $p$-spin dynamics. From top to bottom we show $p=2,3,4$ (with control parameter values $s=0.65, 0.75, 0.8$, respectively). In both left and right boxes we show simulated phase space trajectories and similarity plots constructed by computing the quantity $\mathcal{S}$ discussed in Appendix \ref{sec:app_similarity}, Eq. (\ref{eqn:simi}), over a uniform grid of initial conditions on the unit sphere. Lighter colors correspond to high values of $\mathcal{S}$, which indicate that the simulated trajectories are in good agreement with the ideal mean-field evolution. Likewise, darker colors are associated to small $\mathcal{S}$ and indicate phase space regions which are not accurately simulated. Left box correspond to ensemble size $N=10^4$, and right box to $N=10^6$. For $N = 10^6$ we recover almost the entire phase space with good correlation to the ideal dynamics, except near fixed points and separatrices. Other parameters used in the simulations are $\Delta t = 0.01$, $T_{\rm max} = 3500$, $\mu=25.0$, $n_{\rm cond}=80$, and $n_{\rm sim} = 700$.}\label{fig:ps_simi}
\end{figure*}
Using the protocol with the feedback policy introduced in Sec. \ref{subsec:p_spin_simu} we can simulate arbitrary dynamics of mean-field $p$-spin models. Studying equilibrium phase transitions from dynamical simulations is challenging, since these are associated with the emergence of a new global minimum in the energy function $V(u;s,p)$, which is a static property of the Hamiltonian. However, dynamical changes occurring as a consequence of such static processes are readily accessible in our simulation. Of particular interest are the bifurcation processes and emergence of new fixed points taking place as a function of $s$ for a given value of $p$. These processes are a landmark of the nonlinear character of the mean-field dynamics of the model in Eq. (\ref{eqn:p_spin_hamil}). 

These dynamical processes are seen in the mean-field model and are linked to the radical way in which the phase space changes as a function of the control parameter $s$. For $s<s_{\rm b}$, where $s_{\rm b}$ is the value at which a bifurcation occurs or new extrema appear, the dynamics for any initial condition on the unit sphere is mostly linear, with the Larmor precession trajectories being only slightly deformed by the additional nonlinear term. For values of $s>s_{\rm b}$, major changes in the structure of phase space trajectories occur. In the following we focus on consequences of such changes for the different applications explored with our protocol. 

As a first step we study the degree to which these dynamical changes can be observed with the measurement-based feedback simulator. For this, we take a set $\{\vec{X}_k\}_{k=1,..,n_{\rm cond}}$ of initial conditions on the unit sphere, and evolve them with our scheme according to the map in Eq. (\ref{eqn:explicit_one_step_evo}). With these trajectories we construct the respective phase space portraits. In Fig. \ref{fig:ps_simi} we display these portraits for $p=2,3,4$ with the values $s=0.65, 0.75,0.80$ from top to bottom, respectively. We chose values of $s>s_{\rm b}$ so that major changes in phase space have already taken place. For $N=10^6$ (right box in Fig. \ref{fig:ps_simi}), we see phase spaces with smooth trajectories and we are able to resolve all the fixed points, stable and unstable, and the separatrix line. However, for $N=10^4$, the simulation is not sufficiently deep in the thermodynamic limit to fully reproduce the mean-field dynamics. In this case, the simulated evolution is disturbed by the presence of significant quantum noise, i.e. high projection noise relative to the mean spin, combined with the shot noise which is fixed by optimal measurement strength. This amounts to blurring of the average separation between trajectories which are macroscopically distinguishable. The smoothness of the simulated phase space in this case is greatly reduced. In particular, unstable fixed points and separatrix lines become hard to resolve.

Qualitatively, the phase portraits in Fig. \ref{fig:ps_simi} show good agreement between the QFC simulation and the mean-field phase space, including the deformation of trajectories due to bifurcations and emergence of new extreme points. However, small imperfections can be hard to see by eye at the global scale at which we are looking at the phase spaces. In order to quantitatively assess the quality of the simulated phase spaces, we compute a similarity measure $\mathcal{S}$ between the QFC simulation and the mean-field model (the explicit form of the mean-field map is given in Appendix \ref{sec:app_classical_dpt}). We employ a similarity measure based on the Pearson correlation coefficient~\cite{Lee1988}; its explicit construction is discussed in Appendix \ref{sec:app_similarity}. Its main property is that $\mathcal{S}\in [0,1]$, achieving $\mathcal{S}=1$ for perfect correlation (i.e., if the simulated and mean-field phase spaces are exactly the same), and $\mathcal{S}=0$ for no correlation (i.e., if the simulated and mean-field phase spaces are completely different).

In Fig. \ref{fig:ps_simi} we present point-to-point similarity maps between the mean-field phase spaces and the simulated ones, constructed from computing $\mathcal{S}$ over a uniform grid with $n_{\rm sim}\times n_{\rm sim}$ initial conditions. A few interesting points are manifested from these similarity maps. First, unstable fixed points and their respective separatrix lines are difficult to simulate with a high degree of similarity. Second, and perhaps more striking, trajectories in the vicinity of stable fixed points are also difficult to simulate with a high degree of similarity. This is connected to the fact that, even when working at the optimal value of the measurement resolution $\sigma$, our simulator has a finite resolution power and trajectories in the vicinity of stable fixed points within a range smaller than the resolution are seen as point-like objects. Finally, for initial conditions far from any of these three regions our simulation produces trajectories with an almost perfect similarity (see the right box in Fig. \ref{fig:ps_simi}).\\

The similarity parameter also reveals an important distinction between the degree of success of the simulation scheme for different system sizes. By comparing the left and right similarity plots of Fig. \ref{fig:ps_simi}, we observe that for $N=10^4$ the simulation scheme performs much better for $p=2$ when compared with $p=3$ and $p=4$. This behavior originates from the fact that the latter cases are associated with a proliferation of fixed points after the phase transition occurs, as discussed in Sect. \ref{subsec:p_spin_models}. As a consequence of this, some phase space structures are harder to resolve in the presence of large quantum noise relative to the mean field. This last observation suggests a deeper study of the behavior of $\mathcal{S}$ as we change the system size, which can be used as a probe of the quantum to classical transition, as we will see in Sec. \ref{subsec:quantum_classical}. 

\subsection{Exploring dynamical phase transitions}
\label{subsec:dpt}
\begin{figure}[!t]
 \centering{\includegraphics[width=\linewidth]{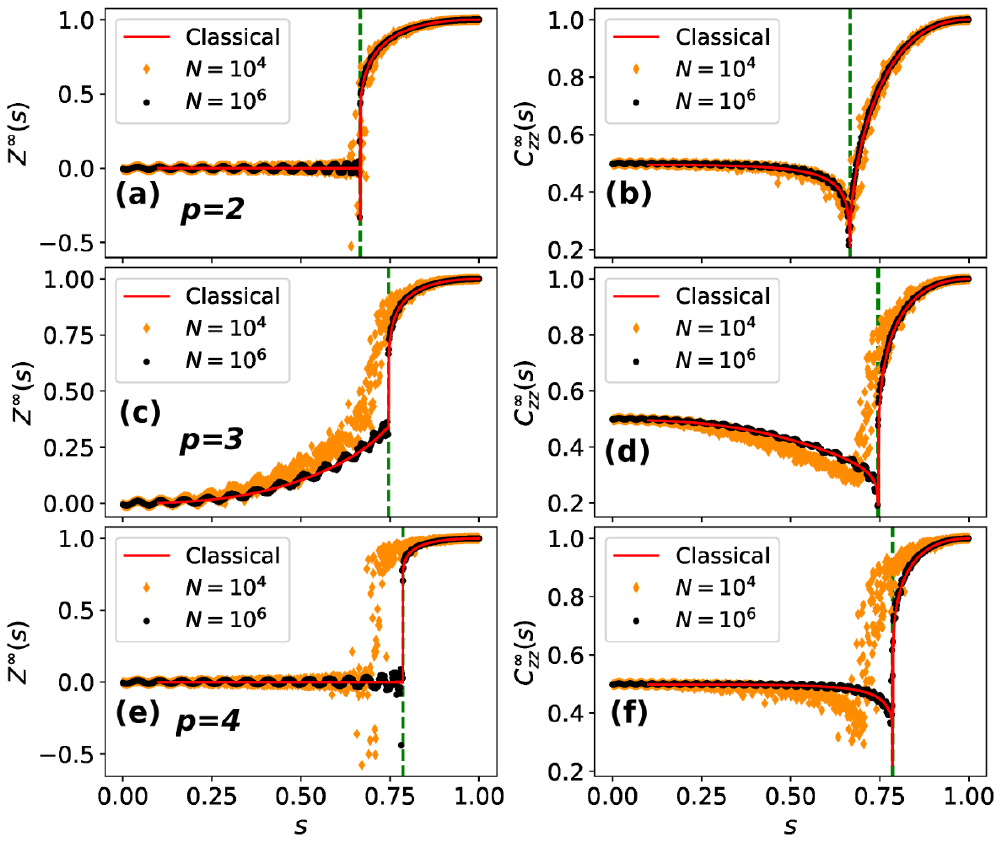}}
\caption{Dynamical phase transition in the simulation of the mean-field $p$-spin model. \textbf{(a,c,e)} Infinite time average order parameter $Z^\infty$, c.f. Eq. (\ref{eqn:dpt_op1}), as a function of parameter $s$. \textbf{(b,d,f)} Infinite time average of the two-body correlation $C_{zz}^\infty$, c.f. Eq. (\ref{eqn:dpt_op2}), as a function of parameter $s$. From top to bottom we display $p=2$ in \textbf{(a,b)}, $p=3$ in \textbf{(c,d)} and $p=4$ in \textbf{(e,f)}. In all figures the symbols show the results obtained using the feedback-based scheme for $N=10^4$ (black dots) and $N=10^6$ (orange diamonds), while the continuous red lines show the values obtained for the classical $p$-spin model described by the flow in Eq. (\ref{eqn:classical_flow}). The dashed green lines indicate the value of the dynamical critical point corresponding to the classical limit. The explicit calculation of the former is given in Appendix \ref{sec:app_classical_dpt}.}\label{fig:dpt}
\end{figure}

As discussed in Sec. \ref{subsec:p_spin_models}, $p$-spin Hamiltonians exhibit equilibrium phase transitions of varying order depending on $p$. Critical behavior can also be found in nonequilibrium properties of quantum systems, leading to \textit{dynamical} quantum phase transitions (DQPTs) \cite{heyl2018}. A characteristic property of DQPTs is the abrupt change of the asymptotic value of the quasi-steady state of an observable which begins out of equilibrium as a function of a parameter in the Hamiltonian \cite{zunkovic2018}. For spin systems, such an observable can be the collective magnetization or a two-body correlation function. Since these quantities are experimentally accessible, DQPTs have attracted much attention as testbeds for near-term quantum simulators, with notable experimental studies including analog quantum simulations of the 1D Ising model with trapped ions \cite{zhang2017_monroe,jurcevic2017} and of the LMG model with superconducting qubits \cite{xu2019}. Recent theoretical works have studied DQPTs in nonintegrable Ising Hamiltonians and their connection to the mean-field limit \cite{zunkovic2018}, and also the relation between different manifestations of DQPTs \cite{sciolla2011,zunkovic2016}.  In the following we present an analysis of DQPTs for $p$-spin models in the mean-field limit, and show how our feedback protocol allows us to access such features of dynamical criticality.\\

In the mean field limit of the $p$-spin model we can build a simple and useful intuition regarding the physical meaning of the dynamical phase transition. Recall that, using the semiclassical energy function, we associated the equilibrium phase transition with the existence of a new global minimum. However, in order for a new global minimum to exist, the semiclassical energy function must have developed new extreme points first. New stable fixed points indicate a major reconfiguration of the trajectories in phase space and are accompanied by new unstable fixed points, which in turn indicate the emergence of a separatrix line. The separatrix marks the boundary between two disconnected regions of phase space, which show different types of regular motion.

In this spirit, the system under study will undergo a phase transition of dynamical character whenever the initial condition finds itself inside a different region of phase space \cite{sciolla2011,zunkovic2018}. As a consequence, the long time average of an order parameter will undergo a major change. Notice that, for systems such as those studied here, the dynamical transition cannot occur without the equilibrium phase transition taking place first. Hence, we expect to find $s_c^{\rm DPT} \geq s_c^{\rm (eq)}$. A detailed explanation of how to compute the mean-field dynamical critical point $s_c^{\rm (DPT)*}$ is given in Appendix \ref{sec:app_classical_dpt}, where we explicitly give the values for $p$-spin models with $p=2,3,4$. 

In Sec. \ref{subsec:phase_spaces} we saw that for an appropriate value of $N$, our measurement-based simulation can capture all the macroscopic changes experienced by the mean-field trajectories, as exhibited in Fig. \ref{fig:ps_simi}. This includes bifurcations and the emergence of the separatrix line. With this capability, we expect that dynamical phase transitions are accessible with our scheme. For the present study we consider two different observables: the long time average of the $z$-magnetization and the two-body correlation function as indicators of the dynamical phase transition. These are given by
\begin{eqnarray}
\label{eqn:dpt_op1}
Z^\infty &=& \lim_{T\to\infty}\frac{1}{T}\int_0^T \frac{\langle \hat{J}_z\rangle_t}{J}dt,  \\
\label{eqn:dpt_op2}
C_{zz}^\infty &=& \lim_{T\to\infty}\frac{1}{T} \int_0^T \frac{\langle\hat{J}_z^2\rangle_t}{J^2} dt,
\end{eqnarray}
where $\langle\hat{O}\rangle_t = \langle\psi(t)|\hat{O}|\psi(t)\rangle$ for any observable $\hat{O}$. Long time averages are good indicators of phase transitions as follows from the theory of dynamical systems and their stability. The details of the trajectories (usually oscillations) around a stable fixed point strongly depend on the initial condition and parameter values. However, time averaged values are usually robust to (small) changes in both the initial condition and model parameters (see, e.g, Chap. 6 of~\cite{Arnold1992}). We expect then, after fixing the initial condition, to see similar or smoothly varying behaviors of $Z^\infty$ and $C_{zz}^\infty$ on each side of the critical point, separated by a sharp transition.   

To explore the dynamical phase transition with our simulator, we prepare an initial spin coherent state along $z$, $|\psi_0\rangle = |J,J\rangle= \lvert \uparrow_z\rangle^{\otimes N}$, and evolve it with our scheme using a fixed value of $s$. After sufficient time of evolution, we approximate the values of $Z^\infty$ and $C_{zz}^\infty$. The procedure is repeated for all values of $s \in [0,1]$. The results of our numerical simulations are shown in Fig. \ref{fig:dpt}, where from top to bottom we display $p=2,3,4$, respectively. We make several observations from these results. First, note that for all values of $p$ the dynamical phase transition is continuous, even in the exact mean-field case (continuous curve in Fig. \ref{fig:dpt}). This is a consequence of the continuous manner in which phase space trajectories are deformed. However, for increasing $p$ the transition becomes sharper. Second, for large $N$ (small projection noise relative to the mean-field), our simulation scheme reproduces almost perfectly the mean-field DPT, including the correct position of the critical point (see black dots in Fig. \ref{fig:dpt}). For a smaller value of $N$ (orange diamonds) our scheme underestimates the value of the critical point for $p>2$, even though the shape of the transition is qualitatively well reproduced. Overall, we observe that the details of the dynamics are harder to reproduce for increasing $p$, in agreement with our analysis of the similarity of the phase space portraits presented in the previous section. We will continue to analyze the behavior of the dynamical phase transition in Sec. \ref{subsec:quantum_classical}.  

\subsection{Spontaneous symmetry breaking}
\label{subsec:symmetry_breaking}
Our protocol allows us to explore aspects of symmetry breaking induced by the action of the measurements (see~\cite{Garcia-Pintos2019} for a related study of measurement induced symmetry breaking in Ising chains).
\begin{figure}[!t]
 \centering{\includegraphics[width=0.48\textwidth]{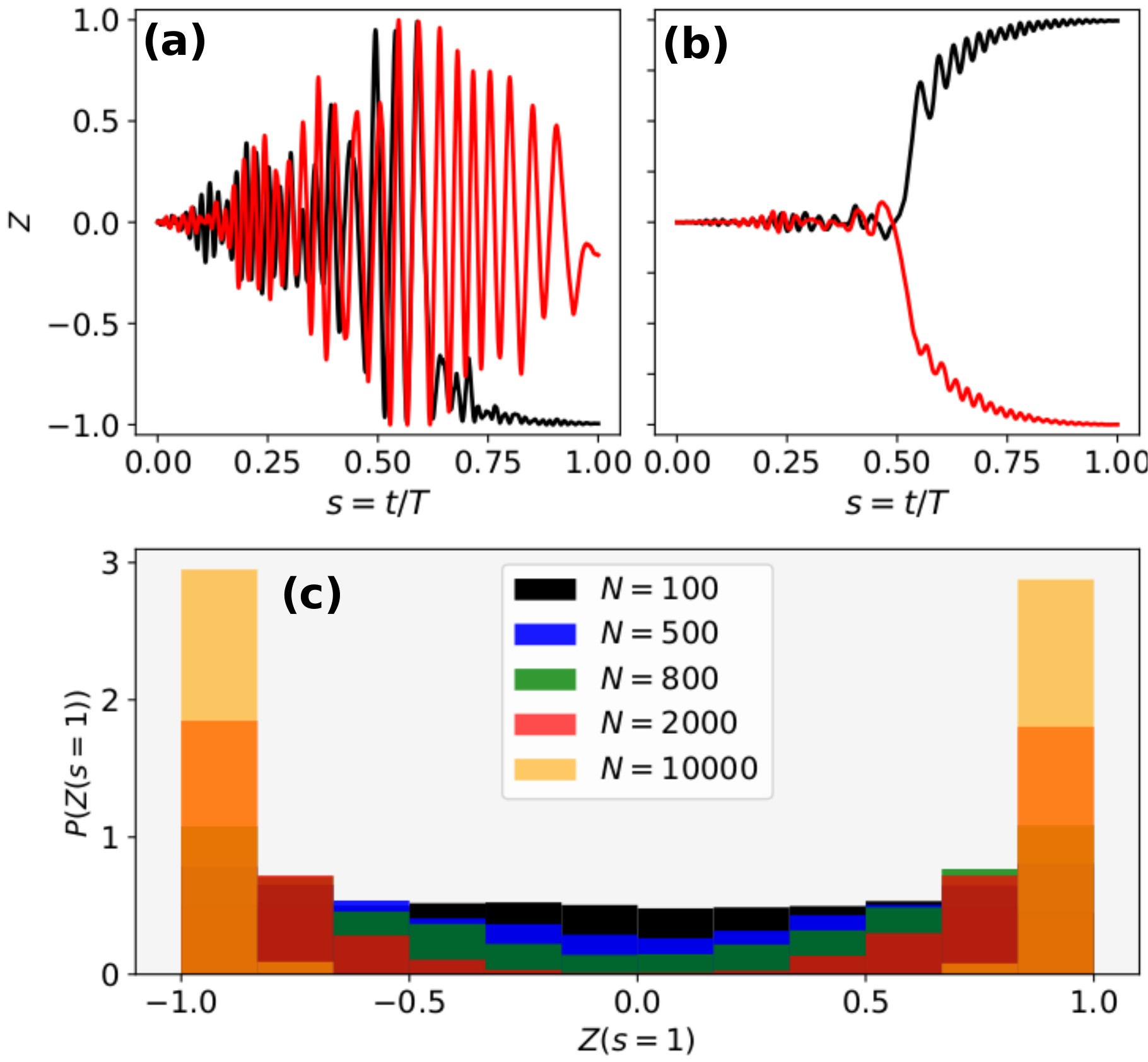}}
\caption{\textbf{(a,b)} Magnetization $Z(t)$ as a function of time during the adiabatic passage, which is described by the parametrization $s=t/T$, where $T$ is the total evolution time. Results are shown for \textbf{(a)} $N = 10^3$ and \textbf{(b)} $N=10^5$. The continuous red and black lines show two different runs of the adiabatic evolution, highlighting the large effect produced by random variations in the simulated adiabatic evolution and in particular of the final values $Z(s=1)$. \textbf{(c)} probability distribution of the values $Z(s=1)$ constructed from $8000$ different runs of the adiabatic evolution. As we decrease the effects of quantum noise relative to the mean-field,  we approach a limit where the statistics of the final values $Z(s=1)$ follow that of a fair coin. This behavior is indicative of the scheme simulating the expected symmetry breaking of the final ground state. Simulation parameters are: $\mu=45.0$, $\Delta t = 10^{-2}$ and $T = 10^4$.} \label{fig:symmetry}
\end{figure}
We focus on the simulation of the $p$-spin model with $p=2$. To study this phenomenon, we consider an adiabatic passage starting from the initial state $|\psi_0\rangle = \ket{\uparrow_y}^{\otimes N}$, the ground state of $-\hat{J}_y$. The adiabatic evolution is generated via $s(t)=t/T$, with $T$ the total passage time. This total time is chosen long enough to guarantee adiabaticity \cite{albash2018} and so the system is expected to reach the ground state of $-\hat{J}_z^2$ at $t=T$, i.e. when $s=1$. This final ground state is an equal-weight superposition of $\lvert\uparrow_z\rangle^{\otimes N}$ and $\lvert\downarrow_z\rangle^{\otimes N}$, and thus one expects that $\langle\hat{J}_z\rangle=0$. In the semiclassical picture, the initial state $(X,Y,Z) = (0,1,0)$ is a fixed point of the flow map in Eq. (\ref{eqn:classical_flow}) and regardless of its stability, this point will be stationary along the adiabatic passage, thus leading to $\langle\hat{J}_z\rangle/J=0$ for all times. However, this point constitutes a set of measure zero, and any imperfection or perturbation will drive the system away from this stationary state in the unstable regime  (that is, for $s > s_c$) \cite{Pilatowsky-Cameo2020}. In the adiabatic evolution simulated by the QFC scheme, it is the backaction induced by the measurement who plays the role of such perturbation, thus generating the symmetry breaking \cite{Garcia-Pintos2019}.

We explore how this measurement-induced symmetry breaking is manifested in the simulation. Note that in this example the protocol simulates a time-dependent Hamiltonian. This is achieved by setting $s\to t_l/T$ in the feedback unitary map of Eq. (\ref{eqn:p_spin_feed_policy}), where $t_l\in[0,T]$, thus realizing a discretized version of $s(t)=t/T$. With this parametrization the simulation proceeds following Eq. (\ref{eqn:explicit_one_step_evo}) where each step has a slightly larger value of $s$, the number of steps follows from the total passage time $T$ and is chosen such that adiabaticity is guaranteed. We show numerical results of different realizations of the protocol in Fig. \ref{fig:symmetry} a,b. The red and black continuous lines represent two different realizations of quantum trajectories simulating the adiabatic evolution with the exact same parameters. In all cases it can be seen that the intrinsic randomness induced by the measurement backaction has a large effect on the final state of the system. In Fig. \ref{fig:symmetry}a we consider a system with $N=10^3$, a value that is sufficiently far from the thermodynamic limit that the effect of projection noise of the initial SCS is relatively large. As a result, the effect of the noise-driven field dominates over the Hamiltonian evolution, and the symmetry breaking is washed out by strong measurement backaction.

On the other hand, for $N=10^5$ (Fig. \ref{fig:symmetry}b), we are sufficiently deep in the thermodynamic limit such that the effect of quantum noise is small compared to the mean-field. As a consequence, the combination of the classical instability and weak measurement backaction is able to break the symmetry and push the system to one of the two final states $\lvert\uparrow_z\rangle^{\otimes N}$ or $\lvert\downarrow_z\rangle^{\otimes N}$. In Fig. \ref{fig:symmetry}c we construct the probability distribution for the expectation value of the spin projection $Z$ at the final time ($t=T$) for different values of $N$. This is done by repeating the simulation many times (8000 in this case) and recording the final state. For small $N$, we obtain an almost uniform distribution in the range $Z\in[-1,1]$ (black histogram), indicating the impossibility of the state to resolve the double well structure of the semiclassical energy function (see Fig. \ref{fig:pseudo_potential}). In this case, the adiabatic evolution essentially realizes a random walk on the sphere. As $N$ increases and the double-well features are resolved, measurement backaction can break the symmetry and we reach the limit of a ``fair coin'' binomial distribution (orange histogram), where all trajectories evolve to either $Z=+1$ or $Z=-1$. The effect of the  finite quantum noise in the symmetry-breaking process can also be used as a probe of the quantum-classical transition, and we will explore this in the next section.
\begin{figure*}[!t]
 \centering{\includegraphics[width=0.98\textwidth]{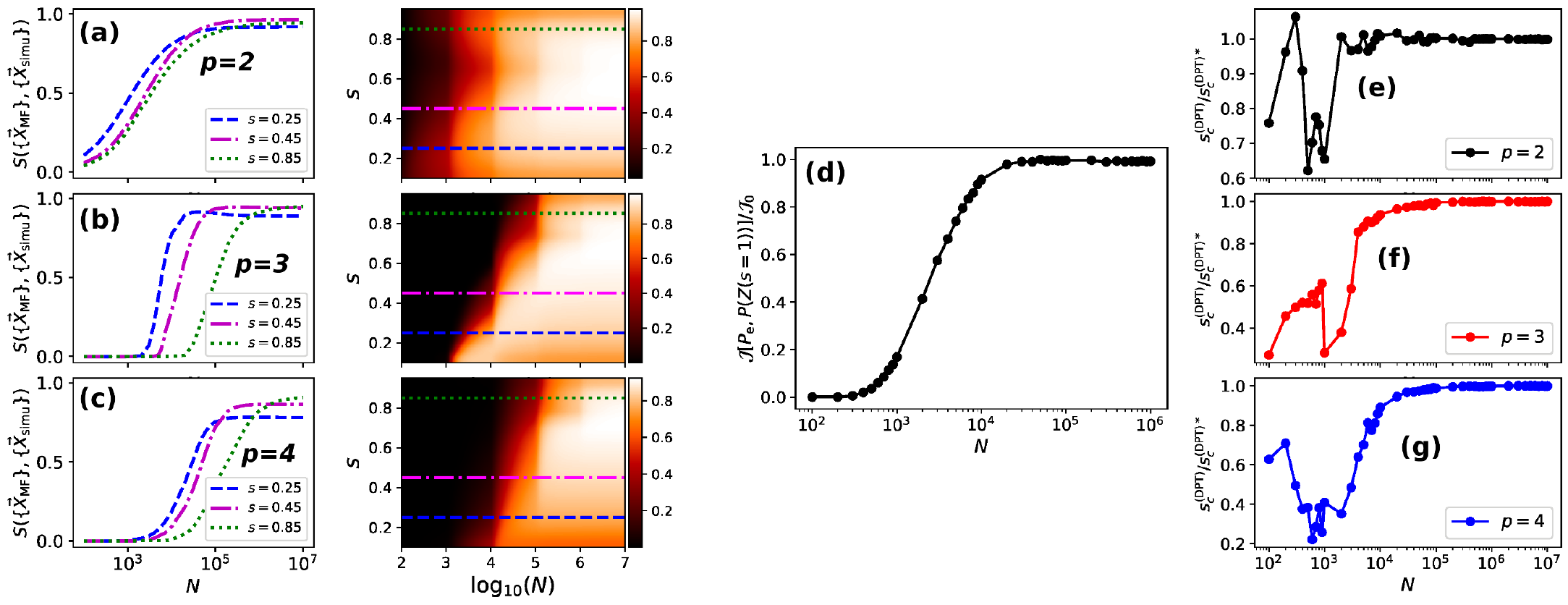}}
\caption{Analysis of the quantum-to-classical transition in the various applications analyzed in this work. \textbf{(a,b,c)} Heat maps show the average similarity as a function of control parameter $s$ and system size $N$ for the cases $p=2,3,4$, top to bottom, respectively. To the left of each map, we plot three sample cross sections corresponding to the three dashed lines on the heat maps. The dashed blue, magenta and green lines correspond to $s=0.25,0.45,0.85$, respectively. \textbf{(d)} Distance of the outcome distributions originally depicted in Fig. \ref{fig:symmetry} \textbf{(c)} to the uniform distribution in the range $Z(s=1)\in[-1,1]$ as a function of the system size $N$, computed using the normalized Jensen-Shannon divergence (see main text for details). \textbf{(e,f,g)} Plots of the dynamical critical point $s_c^{\rm (DPT)}$ obtained through the protocol depicted in Fig. \ref{fig:dpt}, as a function of system size $N$ for $p=2,3,4$, top to bottom respectively. In all cases, values of $s_c^{\rm (DPT)}$ are shown normalized by their classical values, $s_c^{\rm (DPT)*}$ } \label{fig:q_to_c}
\end{figure*}

\subsection{Exploring the quantum-to-classical transition}
\label{subsec:quantum_classical}
The question of how and under what mechanisms a quantum system recovers the appropriate classical behavior, be it regular, chaotic, or critical, has received extensive attention since the formulation of quantum theory. This question and related issues have been explored for closed~\cite{Habib2002,Greenbaum2007,Kapulkin2008,Kumari2018,Pokharel2018,Sakaushi2018,Fink2010}, open~\cite{Habib1998,Nassar2013,Zurek1995,Paz2000} and continuously monitored~\cite{Bhattacharya2000,Bhattacharya2003,Ghose2004,Ghose2003a} quantum systems. The latter is the situation under investigation in the present work. 

More specifically, in this section we explore the emergence of the classical dynamics from the point of view of the three different applications explored in the previous sections. The basic idea here is that in systems of collective spin variables one can introduce an effective Planck constant equal to the reciprocal of the collective spin size, $\hbar_{\rm eff} = 1/J$. Thus, by changing the size of the collective spin we are effectively controlling how deep we are in the classical limit. As seen in the previous section, increasing $J$ reduces the projection noise in the state of the system (relative to the mean spin) and increases the accuracy of the simulation of the mean-field dynamics. This is equivalent to the thermodynamic limit of statistical physics. Here we set out to characterize this behavior in more detail. Particularly, we are interested in understanding how large $N$ should be to accurately reproduce each of the features of criticality studied before. 

First we consider the effects of varying $\hbar_{\rm eff}$ in our ability to reconstruct the mean-field phase spaces. To study this, we use the similarity parameter introduced in Sec. \ref{subsec:phase_spaces} and calculate its phase-space average, c.f. Eq. (\ref{eqn:average_simi}), for a range of values of $s\in[0,1]$ and $N\in[10^2, 10^7]$. Each of these values are displayed as a point in the heat maps of Fig. \ref{fig:q_to_c}a-c, where from top to bottom we have $p=2,3,4$, respectively. Two major features are evident from these plots. First, the region of the space of parameters $(s, N)$ for which our simulator differs substantially from the target model monotonically grows with increasing $p$, as can be seen from the size of the black region in the figures. This behavior is largely due to the increase in the number of fixed points (regardless of their stability). From this result we see that larger $p$ values yield a more difficult model to simulate and pushes the mean-field limit to higher values of $N$. 

A second major feature arises when we look at cross sections of the heat maps (see the dashed lines in Fig. \ref{fig:q_to_c}a-c). In particular we looked at three cross sections for the values $s=0.25,0.45,0.85$ which capture the different phases of the models for the values of $p$ studied. Interestingly, the functional form of all these cross sections for different values of $p$ and $s$ is very similar, hinting at a universal form of the transition to the mean-field limit regardless of the phase and the value of $p$. Finally, as pointed out in the first feature, the only difference between all the cross sections is the position of the inflection point, which shifts to a larger value of $N$ for increasing $s$ and $p$.    

Next, we look at the effects of varying the effective Planck's constant on the statistics of the output of repeated symmetry breaking experiments. We already touched on this analysis in Fig. \ref{fig:symmetry}c, where we showed how the distribution of these outcomes changes between two limiting cases, a uniform distribution in the range $[-1,1]$ at small $N$, and a distribution of two localized peaks at $-1$ and $1$ at large $N$. Furthermore we saw that the distribution changes continuously between these two limiting cases. To quantify this continuous transition we calculated the distance of the outputs distribution to the uniform distribution in $[-1,1]$ using the Jensen-Shannon divergence~\cite{Nielsen2010,Nielsen2019}. This is a symmetrized version of the popular Kullback-Liebler divergence~\cite{Kullback1951}, which satisfies the triangle inequality and thus defines a proper distance in probability space. It is defined as 
\begin{equation}
\label{eqn:jensen_shannon_1}
\mathcal{J}[P_1, P_2] = \frac{1}{2}\left(\mathcal{D}[P_1,P_{\rm m}] + \mathcal{D}[P_2, P_{\rm m}] \right),
\end{equation}
where $\mathcal{D}[P_1, P_2] = -\sum_j P_1^{(j)}\ln\left(\frac{P_1^{(j)}}{P_2^{(j)}}\right)$ with $P_1$, $P_2$ two probability distributions, and $P_{\rm m} = \frac{1}{2}(P_1 + P_2)$. In terms of the Shannon entropy, $S[P] = -\sum_j P_j\ln(P_j)$, Eq. (\ref{eqn:jensen_shannon_1}) can be written in the form
\begin{equation}
\label{eqn:jensen_shannon_2}
\mathcal{J}[P_1, P_2] = S\left[\frac{P_1 + P_2}{2}\right] - \frac{1}{2}\left(S[P_1] + S[P_2]\right).
\end{equation}
Results of our calculation of this quantity are shown in Fig. \ref{fig:q_to_c}d, where the normalization factor $\mathcal{J}_0 = 1/2$ corresponds to the Jensen-Shannon divergence between the uniform distribution and a distribution of two delta functions. Indeed, we observe that the transition between the two limiting distributions is smooth. Additionally the functional form of the Jensen-Shannon divergence when varying $N$ is very similar to that of the cross sections in Fig. \ref{fig:q_to_c}a-c characterizing the similarity of the simulated and ideal phase spaces.


Finally, we study the effects of varying $\hbar_{\rm eff}$ in our simulation of the dynamical phase transition. This is done by an exhaustive numerical study of the position of the critical point as a function of $N$, for $p=2,3,4$. The results are presented in Fig. \ref{fig:q_to_c}e-g, where all the curves are normalized to their respective mean-field critical point (which are reported in Appendix \ref{sec:app_classical_dpt}). Two different behaviors are observed for the cases of $p=2$ and $p>2$. Analogous to the similarity results in Fig. \ref{fig:q_to_c}a-c, the position of the critical point is more resilient for $p=2$ than $p>2$, as seen from the wider plateau at $1$ for $p=2$ in Fig. \ref{fig:q_to_c}d. In addition we see that the behavior for $p=2$ presents two regimes, one in which we recover the correct mean-field dynamical critical point, and one in which our simulation yields a completely different one. On the other hand, for $p>2$ three regimes are observed, one in which the mean-field dynamical critical point is recovered almost with no error, one in which the critical point is smoothly shifted to smaller values, and one in which our simulator cannot yield a physically meaningful value for the critical point. The existence of an intermediate regime indicates a marked difference between the cases $p=2$ and $p>2$. This difference arises because, for the latter, the shape of the $Z^\infty$ and $C_{zz}^\infty$ curves in Fig. \ref{fig:dpt}c-f is mostly maintained with the discontinuity moved to smaller values of $s$. For the former, it comes from the fact that the $Z^\infty$ and $C_{zz}^\infty$ curves in Fig. \ref{fig:dpt}a,b transition from being almost identical for different values of $N$, to completely noisy and not physically relevant curves. We can understand this behaviour from the nature of the birth process of the separatrix line and its vicinity in the mean-field phase space. For $p=2$, the separatrix is generated from the change in stability of a fixed point and thus it is surrounded by an unstable manifold which is fragile in the presence of noise. On the other hand, for $p>2$ the separatrix line appears as a consequence of the appearance of new pairs of stable-unstable fixed points, without a change in stability of the original ones, and thus it is surrounded by stable manifolds. These are more robust to the presence of noise coming from a simulation further away from the mean-field limit. Finally, we also observe that the functional form of the curves in Fig. \ref{fig:q_to_c}f,g presents certain resemblance to those in Fig. \ref{fig:q_to_c}a-d providing more evidence of a very general transition to classical behavior in the proposed simulation scheme.

\section{Summary and outlook}
\label{sec:future}
In this paper we have analyzed in detail and significantly extended upon the method proposed in \cite{munoz2019} for simulating nonlinear dynamics of collective spin systems using quantum measurement and feedback. The method uses unsharp measurements followed by unitary dynamics conditioned on the measurement outcome. Generally, we show that by performing a well-chosen unitary map conditioned by the measurement outcome of an operator $\hat{A}$, one can simulate the dynamics of a Hamiltonian
proportional to $\hat{A}^p$. We focused on collective spin systems and showed that the proposed protocol is particularly suitable to simulate dynamics of a family of models given by $p$-spin Hamiltonians. We demonstrated how different features of these models can in principle be simulated with this scheme. These include phase space structures, spontaneous symmetry breaking and the signatures of dynamical phase transitions. For the latter, we have also obtained novel results that show the emergence of dynamical criticality in the previously unexplored regime of $p>2$. We also presented an extended discussion of the effects of added noise (varying system size $N$), in the different applications explored. The results of this analysis can be seen as a benchmark of the performance of the simulation scheme when the target dynamics is that of the mean-field $p$-spin model, yielding a way of comparing the simulation complexity of different models. Also, by introducing the effective Planck constant $\hbar_{\rm eff} = 1/J$, we can also interpret this analysis as an study of the quantum-to-classical transition. Interestingly, we found unifying features of these transition across different values of the control parameter $s$ and the model parameter $p$ for all the applications considered.

The applications explored in this work and in~\cite{munoz2019} provide evidence of the scope and flexibility of using measurement-based feedback for quantum simulation of Hamiltonian dynamics in the thermodynamic limit, where the Gaussian approximation has validity. Nevertheless, we believe that this simulation scheme is not restricted to mean-field dynamics only and constitutes a platform to investigate dynamical phenomena beyond the Gaussian approximation. This includes, as suggested in~\cite{Lloyd2000}, exploring novel forms of quantum chaos. However, for this method to be useful, one should be able to discriminate the nonlinear effect of the simulated dynamics from the quantum noise. We expect this problem could be tackled using noise characterization techniques which are extensively used in the dynamical systems community \cite{Rosso2007,Ravetti2014,Lacasa2010}. Exploring the domain of purely quantum nonlinear dynamics with the QFC simulation scheme is an exciting avenue for future research.

The $p$-spin models offers additional interesting avenues for future research. One of them is related to different notions of DPTs other than the one presented in this work. It is known that DPTs can also be characterized in terms of the appearance of zeros in the survival probability \cite{heyl2018}. For $p=2$, this notion of DPT and its relation to the transition in the long-time averaged order parameter has been previously studied \cite{zunkovic2018}; the behavior for general $p$ is not known. Also, abrupt changes in the state of a system, as phase transitions, can be characterized using tools from catastrophe theory~\cite{Stewart1982,Poston1996}. Recently, a study of catastrophes for a quantum system of the type of those studied in the present manuscript, with $p=2$, was presented in~\cite{Goldberg2020}. Extending such study for the whole family of $p$-spin models and exploring the consequences of a noisy simulation on the observed catastrophes is another research avenue. In addition, the behavior of the order parameter around the critical point should in principle be described by critical exponents~\cite{Stanley1987}, which have been explored for DQPTs in short-range Ising models~\cite{Heyl2015}. The nature of these exponents for long-range models, and how we can obtain them from our collective spin simulator, remain open problems for future study.

Finally, an important issue in assessing the viability of this protocol in an actual experimental implementation is to study the effects of physical decoherence. In this work we considered an ideal quantum nondemolition measurement, with noise arising solely from the shot noise of the meter and projection noise in the collective spin. Any real implementation will be accompanied by additional imperfections and fundamental decoherence in the system-meter coupling. In Ref. \cite{munoz2019} we studied a simple decoherence model based on the atom-light interface. There we showed that the extraction of Lyapunov exponents from quantum trajectories was still possible even in presence of small decoherence rates. It will be essential to analyze how the features presented in this work can be extracted in a realistic measurement model.

\section*{Acknowledgments}
We thank Daniel Lidar and Tameem Albash for suggesting to us the idea of $p$-spins as a potentially interesting model for such simulations, and Andrew Doherty for his insights on measurement-based feedback control.  This work was supported by NSF grants PHY-1606989, PHY-1630114 and PHY-1912417.


\appendix 
\section{Construction of similarity measure}
\label{sec:app_similarity}
Here we present the details of the similarity measure $\mathcal{S}$ employed in Sec. \ref{subsec:phase_spaces}. In order to compare the classical phase space with that reconstructed from our simulation the similarity between two phase spaces is computed as follows. Let $\{\vec{X}_k\}_{k=1,..,n_{\rm cond}}$ and $\{\vec{X}'_k\}_{k=1,..,n_{\rm cond}}$ be two sets with $n_{\rm cond}$ trajectories corresponding to the mean-field phase space and the simulated phase space, respectively. Each of the trajectories in both sets are generated up to the same final time $T_{\rm max}$, and are obtained from the time evolution of the same set of initial conditions on the unit sphere. Consider a trajectory on each set, say $\vec{X}_{k}$ and $\vec{X}'_k$, belonging to the same initial condition, we quantify their similarity by the product of the Pearson correlation coefficients~\cite{Lee1988} of their three Cartesian components extended in time, 
\begin{equation}
\label{eqn:simi}
 \mathcal{S}(\vec{X}_k, \vec{X}'_k) = \left| {\rm cor}(\tilde{X}_k, \tilde{X}'_k){\rm cor}(\tilde{Y}_k, \tilde{Y}'_k){\rm cor}(\tilde{Z}_k, \tilde{Z}'_k)\right|,
\end{equation}
where $\tilde{X}_k = (X_k^{(1)}, 
X_k^{(2)},...,X_k^{(n_t)})$ with $n_t$ the number of time bins in the interval $[0, T_{\rm max}]$, is the vector of all the $X$ components from initial 
time to final time for the $k$-th trajectory. The Pearson correlation coefficient is given by 
\begin{equation}
 \label{eqn:pearson}
 {\rm cor}(A,B) = \frac{{\rm cov(A, B)}}{\sqrt{{\rm var}(A){\rm var}(b)}},
\end{equation}
with ${\rm cov}(A,B)$ the covariance between vectors $A$ and $B$ of same length, and ${\rm var}(A)$ the variance of vector $A$. Notice that Eq. (\ref{eqn:pearson}) gives $1$ for perfect correlation between $A$ and $B$, $-1$ for perfect anti-correlation and $0$ in absence of correlations. Note that, in order to have a similarity measure yielding a value strictly in the interval $[0,1]$, we define $\mathcal{S}$ as the absolute value of the product of Pearson correlation coefficients, as in Eq. (\ref{eqn:simi}). However, it is important to point out that for this application only very small negative covariances are found.  

The conditioned evolution over a single time-series of measurement outcomes maps pure states into pure states. Thus, in the mean-field limit, trajectories will always remain close to the surface of the unit sphere. We can then express these trajectories in angular coordinates and define $\mathcal{S}_{\rm ang}$ as 
\begin{equation}
\label{eqn:simi_ang}
\mathcal{S}_{\rm ang}(\vec{X}_k, \vec{X}'_k) = \left|{\rm cor}(\tilde{\theta}_k, \tilde{\theta}'_k){\rm cor}(\tilde{\phi}_k, \tilde{\phi}'_k)\right|,
\end{equation}
where $\tilde{\theta}_k$ is the time ordered vector of the $\theta$ coordinates of the $k'$-th trajectory, and $\tilde{\phi}_k$ that of the $\phi$ coordinates.

The detailed study of the phase space similarities presented in Sec. \ref{subsec:quantum_classical} as a function of the system size $N$, uses a overall similarity score given to the whole phase space, as obtained from a set of initial conditions uniformly distributed over the unit sphere. We construct this phase space similarity score by taking the average of $\mathcal{S}$ over the $n_{\rm sim}\times n_{\rm sim}$ grid of initial conditions
\begin{multline}
 \label{eqn:average_simi}
 \overline{\mathcal{S}} = \frac{1}{n_{\rm 
sim}^2}\sum_{k=1}^{n_{\rm sim}^2}\mathcal{S} \left( \{\vec{X}_k\}_k, 
\{\vec{X}'_k\}_k \right) \\
= \frac{1}{n_{\rm sim}^2}\sum_{k=1}^{n_{\rm sim}^2}\left|{\rm cor}(\tilde{X}_k, \tilde{X}'_k){\rm 
cor}(\tilde{Y}_k, \tilde{Y}'_k){\rm cor}(\tilde{Z}_k, \tilde{Z}'_k)\right|.
\end{multline}
For this quantity a value of $\overline{\mathcal{S}} = 1$ tells 
that the two phase spaces are identical, and $\overline{\mathcal{S}} 
= 0$ tells the two phase spaces are completely different.

\section{Calculation of the DPT critical values}
\label{sec:app_classical_dpt}
In Sec. \ref{subsec:dpt} and Sec. \ref{subsec:quantum_classical}, we presented simulations of the dynamical phase transitions, that compare the dynamical critical point obtained within the simulation with that obtained from the mean-field model. Here we show how to compute the values of the dynamical critical point in the later case. 

Our DPT protocol follows the evolution of the initial condition $|\psi_0\rangle = \lvert\uparrow_z\rangle^{\otimes N} = |J,J\rangle = |\theta=0,\phi=0\rangle$, which in the mean-field limit is given by the vector $(X,Y,Z) = (0,0,1)$. As explained in the main text, the DPT is characterized by  $Z^\infty$ and $C_{zz}^\infty$ showing sharply different behaviors whether the initial condition belongs to one of two separated regions of phase space. Thus the dynamical critical point $s_c^{\rm (DPT)}$, is given by value of $s$ at which the initial condition lies in the separatrix line.

Conservation of energy allows us to obtain $s_c^{\rm (DPT)}$ by finding the value of $s$ which makes the energy of the initial condition equal to that of the separatrix point. That is, $s=s_c^{\rm (DPT)}$ leads to the following equality,
\begin{equation}
\label{eqn:criti_dpt}
V(Z_{\rm sp}; s_c^{\rm (DPT)}, p) = V(Z_0; s_c^{\rm (DPT)}, p),
\end{equation}
where $Z_{\rm sp}$ is the $z$-component of the separatrix point and $Z_0$ is the $z$-component of the initial condition. For $p=2$ the separatrix line appears due to the change in stability of the fixed point at $(X,Y,Z)=(0,1,0)$, then Eq. (\ref{eqn:criti_dpt}) takes  the form $-(1-s) = \frac{s}{2}$ which has solution $s_c^{\rm (DPT)} = 2/3$. For $p>2$ the separatrix line appears as a division between old and new stable fixed points and is not a consequence of a change in stability. Thus, finding the explicit form of $s_c^{\rm (DPT)}$ is more involved than for $p=2$. In particular, when $p=3$, the $z$-component of the separatrix point has the form
\begin{equation}
Z_{\rm sp} = \sqrt{\frac{1}{2} - \frac{1}{2}\sqrt{1-4\left(\frac{1-s}{s}\right)^2}},
\end{equation}
and the numerical solution of Eq. (\ref{eqn:criti_dpt}) yields $s_c^{\rm (DPT)}\approx0.745921$. The expression for the $z$-component of the separatrix point for $p=4$ can easily be found numerically. With this expression at hand one can solve Eq. (\ref{eqn:criti_dpt}), from where we find $s_c^{\rm (DPT)}\approx 0.786074$. Note that a simpler estimate of the critical point can be employed and gives quite accurate results. For $p>2$ the energies of the point $(X,Y,Z)=(0,1,0)$ and the separatrix point are not so different, thus one can use that point in the right hand side of Eq. (\ref{eqn:criti_dpt}). By doing so we find the values $s_c^{\rm (DPT)} = \frac{2}{3}, \frac{3}{4}, \frac{4}{5}$ for $p=2,3,4$, respectively, values which are fairly close to the exact ones.

Identifying the separatrix line can be done by looking at its stability with the tangent map of the flow defining the time evolution of the classical equations of motion. This flow is given by
\begin{eqnarray}
\label{eqn:classical_flow}
\frac{dX}{dt} &=& -(1-s)Z + sZ^{p-1}Y, \nonumber \\
\frac{dY}{dt} &=& -sZ^{p-1}X, \\
\frac{dZ}{dt} &=& (1-s)X, \nonumber
\end{eqnarray}
where the equations are obtained from the mean-field limit of the Heisenberg evolution of $\hat{J}_\gamma$. The tangent map of this set of equations is given by
\begin{equation}
\mathcal{M}(\vec{X}) = \begin{pmatrix}
    0 && sZ^{p-1} && -(1-s)+s(p-1)Z^{p-2}Y \\
    -sZ^{p-1} && 0 && 0 \\
    1-s && 0 && 0
    \end{pmatrix},
\end{equation}
where we have used the fact that any fixed point of the flow in Eq. (\ref{eqn:classical_flow}) has a vanishing $x$-component.

\bibliography{feedback_long}

\end{document}